# No Universal Winner in Model Selection: Regression versus Representational Similarity Analysis

Chuanji Gao[a,b,c*], Gang Chen[d], Svetlana V. Shinkareva[e, f*] and Rutvik H. Desai[e, f*]

[a] School of Psychology, Nanjing Normal University, Nanjing, China
[b] Taizhou Fifth People's Hospital, China
[c] Adolescent Education and Intelligence Support Lab of Nanjing Normal University, Laboratory of Philosophy and Social Sciences at Universities in Jiangsu Province, Nanjing, China
[d] Scientific and Statistical Computing Core, National Institute of Mental Health, Bethesda, MD, United States
[e] Department of Psychology, University of South Carolina, Columbia 29208, USA
[f] Institute for Mind and Brain, University of South Carolina, Columbia 29208, USA

*Correspondence should be addressed to Chuanji Gao, chuanji.gao@njnu.edu.cn; Svetlana V. Shinkareva, shinkareva@sc.edu and Rutvik H. Desai, rutvik@sc.edu.

*Acknowledgements*: We thank Dexin Shi for the helpful discussions. Artificial intelligence (AI) tools were used to improve the language and clarity of the manuscript. All AI-assisted revisions were carefully reviewed and approved by the authors, who take full responsibility for the final content.
*Author contributions*
Chuanji Gao: Conceptualization; Methodology; Software; Formal analysis; Investigation; Data curation; Writing-original draft; Visualization; Funding acquisition.
Gang Chen: Methodology; Software; Investigation; Writing-review & Editing.
Svetlana V. Shinkareva: Conceptualization; Investigation; Writing-review & Editing.
Rutvik H. Desai: Conceptualization; Investigation; Funding acquisition; Writing-review & Editing.
*Declaration of Conflicting Interests*: None.
*Funding*: The work was supported by National Natural Science Foundation of China (32300863), the Open Research Fund of the Joint Research Institute for Brain Science and Mental Health, Nanjing Normal University and Taizhou Fifth People's Hospital (NKX2026Z09), and NIH/NIDCD R01DC017162. G.C. was supported by the NIMH Intramural Research Program (ZICMH002888) of NIH/HHS, USA.
*Ethics*: This study used simulated data and previously published item-level psycholinguistic measures from SCOPE, including lexical-decision responses from the English Lexicon Project. No new human participants were recruited. SCOPE is a metabase that compiles existing datasets and did not itself collect the human responses analyzed here.
*Emails*: Chuanji Gao, chuanji.gao@njnu.edu.cn; Chen Gang, gangchen@mail.nih.gov; Svetlana V. Shinkareva, shinkareva@sc.edu and Rutvik H. Desai, rutvik@sc.edu.
*Data availability:* Data are available at Github (https://github.com/chuanjigao-sys/REGRESSION-AND-RSA-IN-MODEL-SELECTION).
*Code availability:* The custom code for data analysis is available at Github (https://github.com/chuanjigao-sys/REGRESSION-AND-RSA-IN-MODEL-SELECTION).

## Abstract

Selecting among competing computational models is central to cognitive psychology and neuroscience, yet the relative model-selection performance of regression and representational similarity analysis (RSA) remains unclear. We compared these approaches using Monte Carlo simulations and empirical data. Univariate simulations compared generating and rival feature models against the same response under direct-linear and geometry-based mappings. Under direct-linear generation, regression outperformed correlation-distance RSA/Spearman but not RSA/Pearson. Under geometry-linear generation, neither approach showed a significant advantage, whereas correlation-distance RSA outperformed regression under a sigmoid readout, although this advantage did not extend to Euclidean-distance RSA. Feature dimensionality, covariance, and PCA preprocessing further influenced model-selection accuracy. In multivoxel simulations, pattern component modeling (PCM) showed superior performance under spatially correlated noise, while crossnobis improved selection relative to squared Euclidean RSA under the same condition. Feature reweighting improved held-out representational correspondence for both generating and rival models without consistently improving model selection. Separate signal-strength comparisons yielded different method rankings. In empirical data from 4,026 words, all methods preferred frequency over affect features, but small-sample preference stability depended on distance measures and preprocessing. Together, these findings demonstrate that model-selection performance depends on stimulus-response mappings, representational geometry, measurement noise, and inferential objectives.

## Introduction

In psychological and neuroscience research, understanding how computational features of external stimuli relate to behavioral and neural responses is a common methodological challenge. Researchers typically measure responses to a set of stimuli, such as reaction times, behavioral ratings, or multivoxel fMRI activity patterns, while using computational models to characterize the features of those stimuli. For example, word stimuli can be represented by frequency measures derived from different language corpora or by affective dimensions such as valence, arousal, and dominance. A common objective is to determine how well these feature models explain observed responses. More critically, researchers often need to compare competing models to determine which better accounts for the same responses. The ability to accurately distinguish between competing models is central to research in psychology and neuroscience.

Multiple regression is a widely used approach for evaluating the relationship between stimulus features and observed responses. As a first-order approach, regression directly models responses as a function of stimulus features by estimating feature weights that minimize prediction errors (Tukey, 1977; Maxwell et al., 2018). Model performance is commonly evaluated using the coefficient of determination ($R^2$), with adjusted $R^2$ accounting for the number of predictors (Fox, 2015). This approach has been widely applied in psycholinguistic research to evaluate the predictive contributions of different stimulus properties (Cortese & Khanna, 2022; Snefjella & Kuperman, 2016). For example, Brysbaert and New (2009) compared word-frequency measures derived from different sources and found that subtitle-based frequencies better predicted lexical-decision times than frequencies based on written texts. This approach is also widely used in neuroscience, particularly in voxel-wise encoding models that predict neural activity from stimulus features. For example, feature-based encoding models have been used to predict visual cortical responses from image features (Kay et al., 2008) and cortical responses to natural speech from semantic features (Huth et al., 2016). By directly estimating feature-to-response weights, regression can effectively characterize stimulus-response relationships when they are well approximated by a linear mapping. However, direct feature-to-response modeling is not always the most suitable approach, particularly when the scientific question concerns relationships among stimulus representations rather than the prediction of individual response channels.

Representational similarity analysis (RSA) provides a complementary approach by comparing the representational geometry of computational models with that of observed responses (Edelman, 1998; Kriegeskorte, Mur, & Bandettini, 2008; Kriegeskorte & Kievit, 2013; Nili et al., 2014; Popal et al., 2019; Xie et al., 2025). Rather than directly estimating the mapping from stimulus features to responses, RSA constructs representational dissimilarity matrices (RDMs) that characterize pairwise relationships among stimuli in feature and response spaces. Model-response correspondence is then evaluated by comparing these dissimilarity structures. This approach is particularly useful when direct correspondence between representational dimensions is unavailable, such as when comparing different brain regions, measurement modalities, or species (Kriegeskorte, Mur, Ruff, et al., 2008; Tsantani et al., 2019; Cichy & Oliva, 2020). RSA also provides a natural framework for analyzing behavioral similarity judgments (Groen et al., 2018; Riberto et al., 2022).

Despite their widespread use, regression and RSA differ in several important respects that may influence model selection. Regression estimates feature weights to optimize correspondence with observed responses, whereas conventional fixed RSA evaluates a predefined representational geometry without fitting feature weights. Moreover, different RSA distance measures, such as correlation and Euclidean distances, preserve different properties of stimulus representations. Consequently, differences in model-selection performance cannot necessarily be attributed to the distinction between first-order and second-order representations alone (Diedrichsen & Kriegeskorte, 2017). Importantly, the magnitude of model-response correspondence and the ability to distinguish competing models are also different quantities. Relatively low RSA correlations, as sometimes observed in empirical studies (Guo et al., 2023; Tsantani et al., 2019), do not necessarily imply poor discrimination between competing feature models. Thus, a central methodological question is under what conditions regression and RSA differ in their ability to reliably identify the better model.

One potentially important factor is how stimulus features are mapped onto observed responses. When responses arise directly from a linear combination of stimulus features, regression is closely aligned with the underlying generating process. However, behavioral and neural responses may also arise from transformations of the representational geometry of stimuli, rather than from direct linear combinations of their original features. For example, exemplar-based categorization models propose that classification decisions depend on the similarity of a stimulus to stored category exemplars within a multidimensional representational space (Nosofsky, 1986). This suggests that stimulus features may influence observed responses through their representational geometry, providing a theoretical motivation for examining geometry-dependent stimulus-response mappings.

Importantly, geometry-dependent response generation does not necessarily imply an advantage for RSA. Responses derived from representational geometry may still be well approximated by a linear combination of the stimuli features. Moreover, the relative performance of regression and RSA may depend on how the underlying geometry is transformed into observed responses and on how representational distances are defined. For example, a linear readout and a nonlinear readout of the same latent representation may produce different relationships between stimulus features and responses. Whether these differences favor regression or RSA remains an empirical question. Furthermore, the relative performance of these methods may depend on sample size, measurement noise, feature dimensionality, and feature covariance, as well as on preprocessing transformations that alter the geometry of the feature space. Systematically examining these factors can therefore help clarify the conditions under which different analytical approaches provide more reliable discrimination between competing feature models.

Additional considerations arise when model selection involves multivoxel activity patterns. Beyond regression and RSA, pattern component modeling (PCM) provides a covariance-based approach for evaluating whether the representational structure predicted by a computational model is consistent with observed multivoxel activity patterns (Diedrichsen et al., 2011, 2018; Diedrichsen & Kriegeskorte, 2017). The performance of representational methods may also depend on how measurement noise is handled. For example, cross-validated Mahalanobis distances provide a noise-aware approach for estimating representational dissimilarities from repeated neural measurements (Nili et al., 2014; Walther et al., 2016). Another important consideration is whether feature contributions are fixed or learned from the data. Previous studies have shown that feature reweighting can improve correspondence between computational models and neural representations (Jozwik et al., 2016; Kaniuth & Hebart, 2022; Konkle & Alvarez, 2022; Conwell et al., 2024). However, improved representational correspondence does not necessarily imply greater accuracy in distinguishing competing models. This distinction is particularly important when candidate models share response-relevant features, because greater flexibility in fitting feature weights may improve the correspondence of both models without increasing their separability. These considerations highlight the importance of distinguishing representational correspondence from model-selection accuracy.

To address these questions, the present study systematically compared regression, RSA, and PCM using controlled simulations and empirical data. We first evaluated model-selection accuracy in univariate simulations in which two equally sized, competing feature models were evaluated against the same response. We considered two data-generating schemes: a direct-linear mapping from stimulus features to responses and a geometry-to-first-order mapping with either a linear or sigmoid readout. We examined how model-selection accuracy varied with sample size, measurement noise, and the number of replaced features, followed by additional controls of feature dimensionality, covariance, and PCA preprocessing. In a separate analysis, we assessed sensitivity to differences in signal strength using independently generated datasets. We then extended the model-selection framework to simulated multivoxel responses, comparing regression, PCM, and RSA with different distance estimators, including crossnobis. We further evaluated whether learning feature-to-voxel weights improved held-out representational correspondence and model-selection accuracy. Finally, we applied regression and RSA to an empirical lexical-decision dataset to compare frequency and affect feature models and examine the stability of model preference across sample sizes and preprocessing conditions. Together, these analyses aimed to determine how different assumptions about stimulus-response mappings, feature-space structure, model fitting, and measurement noise influence the ability of analytical approaches to distinguish competing computational models. By separating known-truth model selection from signal-strength sensitivity and representational correspondence, this study also sought to clarify whether better model fit necessarily translates into more accurate model selection.

# Method

We first compared regression and representational similarity analysis (RSA) using simulated univariate responses, then extended the comparison to simulated fMRI data, and finally applied the methods to an empirical behavioral dataset. Importantly, we distinguished model selection, in which competing feature models were evaluated against the same response, from signal-strength sensitivity, in which independently generated datasets with different signal strengths were compared. **Figure 1** provides an overview of the data-generating procedures and analytical framework. Each condition comprised 1000 independent replications, with the replication serving as the unit of statistical inference.

For each simulation, X denotes the n-by-p stimulus-feature matrix. Y denotes either a univariate response vector of length n or an n-by-V matrix of multivoxel responses, where V is the number of voxels. In the model-selection simulations, we compared a generating model, from which the response was constructed, with a rival model containing the same number of features. Both candidate models were evaluated against the same observed response within a replication.

Model-selection accuracy and signal-strength identification rates were calculated as the proportions of successful selections across replications, with 95% Wilson confidence intervals. Differences in these rates between methods were evaluated using two-sided exact McNemar tests, with confidence intervals obtained from 10,000 paired bootstrap resamples. For model selection, we additionally calculated paired-samples Cohen's d, which is the mean score difference between the generating and rival models standardized by the standard deviation of within-replication score differences. For the independent-dataset sensitivity analysis, we instead used independent-samples Cohen's d, defined as the difference between the high- and low-signal group means divided by their pooled standard deviation. Comparisons of held-out representational fit between fixed RSA and voxel-encoding RSA (veRSA) additionally used two-sided paired *t* tests on replication-level fold means. Multiple comparisons were controlled using the Holm correction within prespecified test families, with separate corrections for primary analyses, calibration controls, and supplementary or exploratory analyses.

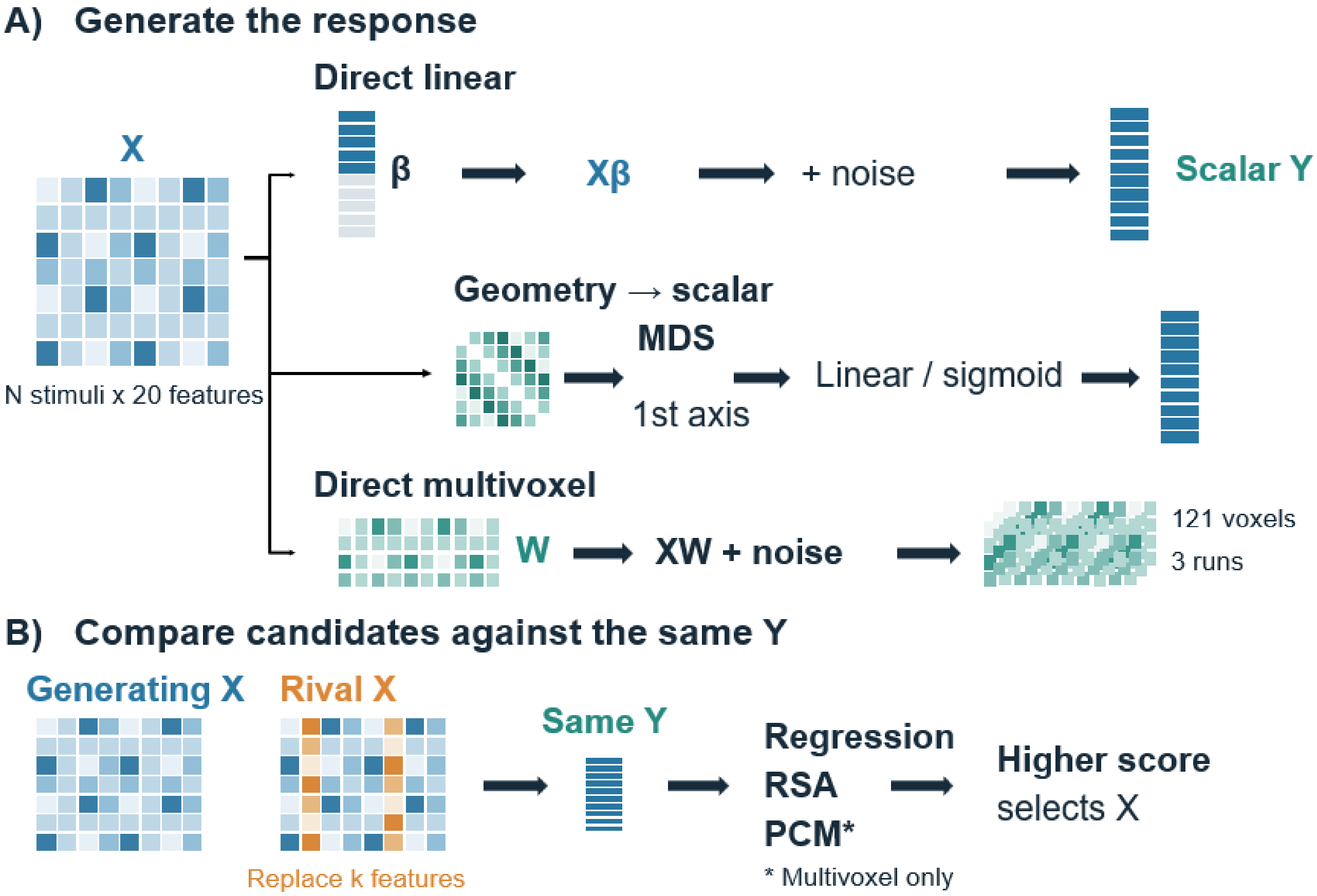

**Figure 1. Data generation and comparison of competing feature models.** A) Scalar responses are generated directly from Xβ, or from the first coordinate of an MDS embedding of the feature RDM followed by a linear or bounded sigmoid readout. Multivoxel responses are generated from XW with feature-specific spatial maps and three noisy runs. B) In each model-selection replication, the generating and rival candidates have equal feature counts and explain the identical response. The rival replaces k columns by conditional Gaussian innovations. Each method selects the candidate with the higher score; selection accuracy is the proportion favoring the generating candidate across replications. PCM is used only for multivoxel data.

## Univariate model selection

The primary univariate simulations evaluated the ability of regression and RSA to identify the generating model among two competing feature models. In each replication, a single response Y was generated from a feature matrix X containing 20 features, and both the generating and rival models were evaluated against the same Y. The rival model was constructed by replacing k features of the generating model while retaining the remaining features. The reference condition included 200 stimuli (n = 200), five replaced features (k = 5), and a noise standard deviation (SD) of 5. We examined the effects of sample size (n = 100, 200, 300, 400, and 500), the number of replaced features (k = 1, 2, 5, and 10), and noise level (SD = 5, 10, and 15).

**Data generation.** For each replication, stimulus feature vectors were sampled from a zero-mean multivariate Gaussian distribution with a specified covariance structure. The 20 features were divided into two blocks of ten features, with within-block correlations of 0.2 and cross-block correlations sampled uniformly from 0 to 0.1. We considered two data-generating schemes. In the direct-linear scheme, the first ten features had regression coefficients of 0.5, while the remaining ten had coefficients of zero, with the intercept fixed at 1. The response was generated by summing the weighted feature values and adding Gaussian noise with the specified SD. In the geometry-to-first-order scheme, all 20 features contributed to the response. We first standardized each feature across stimuli and constructed a representational dissimilarity matrix (RDM) using correlation distance (1 − r), where r is the Pearson correlation between the feature vectors of two stimuli. Classical multidimensional scaling (MDS) was then applied to the RDM to extract its first coordinate, u, which was standardized across stimuli. The response signal was generated using either a linear readout (5u) or a bounded sigmoid readout, defined as $16\left(\frac{1}{1+\exp\left(-\frac{z}{1+\frac{|z|}{3}}\right)}-\frac{1}{2}\right)$, where z = 5u. Gaussian noise was then added to obtain the observed response.

**Construction of the rival model.** For each replication, the rival model was constructed by replacing k columns of X independently of Y. Replacement values were drawn from a conditional multivariate Gaussian distribution, with their mean and covariance determined by the original feature-covariance matrix and the retained features. This design preserved the intended feature covariance structure while allowing the rival model to capture response-relevant information through correlated features. Under the direct-linear scheme, only the ten features with nonzero coefficients were eligible for replacement, whereas under the geometry-to-first-order scheme, any of the 20 features could be replaced.

**Model evaluation.** Both regression and RSA were applied to each pair of candidate models under identical responses and scoring procedures across data-generating schemes. Before analysis, all feature columns and the response vectors were standardized across stimuli. For regression, we fitted an ordinary least squares model with an intercept, quantifying performance using adjusted $R^2$. For RSA, we constructed an RDM for each candidate feature matrix using either correlation distance or Euclidean distance. The response RDM was computed using absolute differences between scalar response values. We then vectorized the lower off-diagonal triangles of the feature and response RDMs and quantified their correspondence using either Pearson or Spearman correlation. Thus, four RSA implementations were evaluated, combining two distance measures with two correlation metrics. Unlike regression, these fixed-RSA implementations did not estimate feature weights from Y. For each method, model selection favored the candidate exhibiting the higher evaluation metric.

## Univariate controls of dimensionality, covariance and preprocessing

We conducted three additional control analyses to examine how feature dimensionality, covariance structure, and preprocessing influenced model-selection accuracy. All controls followed the same model-selection procedures described above, with the reference condition set to n = 200, k = 5, and noise SD = 5.

**Feature dimensionality.** To examine how adding irrelevant features affects model-selection accuracy without altering the underlying stimulus-response relationship, we restricted this control to the direct-linear scheme. We increased the number of features from 20 to 40 and 60 by adding independent Gaussian features with zero coefficients to both candidate models. The original features, generated response, and measurement noise were held constant across conditions, allowing us to assess the effect of increasing feature dimensionality while keeping the underlying signal unchanged.

**Feature covariance.** To examine how correlations among features influenced model-selection accuracy, we varied the correlations among the second block of ten features (features 11-20) across 0, 0.2, 0.4, and 0.8, leaving the covariance structure of the remaining features unchanged. In the direct-linear scheme, these ten features did not contribute to the response, allowing us to manipulate their correlations without altering Y. In contrast, in the geometry-to-first-order scheme, all 20 features contributed to the representational geometry used to generate Y. Consequently, changing feature correlations also changed the underlying geometry and the resulting response.

**PCA preprocessing.** To examine the effects of feature preprocessing, we compared model-selection performance using the original standardized features, PCA-rotated features, and PCA-whitened features. PCA was performed separately for each candidate model without using Y. The PCA rotation retained all 20 components unscaled, whereas PCA whitening further standardized the nonzero principal component scores to unit variance. PCA rotation preserves the column space used by regression and the Euclidean distances between stimuli, but can alter correlation distances because they depend on the coordinate system. PCA whitening additionally changes Euclidean distances by rescaling the principal components, while preserving the regression model's fitted values. Thus, these transformations allowed us to examine how changes in feature representation influenced RSA independently of the information available to regression.

## Sensitivity to differences in signal strength

To examine how regression and RSA detect differences in the strength of stimulus-response relationships, we compared datasets generated with smaller ($\beta = 0.4$) and larger ($\beta = 0.5$) effects. Unlike the model-selection simulations, which compared competing feature models against the same response, this analysis compared independently generated datasets, each with its own feature matrix X, response Y, and measurement noise. The number of features was fixed at 20, with the same covariance structure described above. We varied sample size (n = 100, 200, 300, 400, and 500) and noise level (SD = 5, 10, and 15), with SD = 5 used for the primary sample-size comparison.

**Data generation.** In the direct-linear scheme, the first ten features had coefficients of either 0.4 or 0.5, while the remaining ten had coefficients of zero. In the geometry-to-first-order scheme, the same manipulation was applied to the latent geometric signal. Specifically, the response was generated using either a linear readout or the sigmoid readout described above.

**Model evaluation.** Each dataset was evaluated using regression (adjusted $R^2$) and the four RSA implementations described above. For each method, we compared the score from the larger-effect dataset ($\beta = 0.5$) with that from the smaller-effect dataset ($\beta = 0.4$). A successful identification was recorded when the larger-effect dataset received a higher score. We calculated the proportion of successful identifications and summarized the score distributions using their means, standard deviations, and independent-samples Cohen's d. These measures reflect sensitivity to differences in signal strength rather than accuracy in selecting competing feature models for the same response.

## Multivoxel model selection and distance estimators

To examine model-selection performance with multivoxel responses, we extended the direct-linear simulation framework to fMRI-like data. Each replication included 20 stimulus features and responses from 121 voxels measured across three runs. As in the univariate simulations, a generating model and a rival model were evaluated against the same responses. The rival model was constructed by replacing one of the ten active features (k = 1), using the same conditional Gaussian procedure described above. We varied sample size (n = 100, 200, 300, 400, and 500) and noise level (SD = 5 and 20), and included identical-candidate and pure-noise controls. An additional exploratory analysis examined the replacement of five active features (k = 5) at n = 200.

**Data generation.** Stimulus features were generated using the same Gaussian covariance structure as in the univariate simulations. To generate multivoxel responses, we constructed a $20 \times 121$ weight matrix W, assigning distinct spatial weight patterns to the ten active features, each containing positive and negative weights scaled to a root-mean-square amplitude of 0.5. The remaining ten features had zero weights. The resulting noise-free response was XW, producing a signal matrix with a rank of ten. This signal was measured across three independent runs, each with added Gaussian noise. Voxels were arranged on an $11 \times 11$ grid, and we considered two noise structures: spatially independent noise and spatially correlated noise, in which correlations between voxels decreased with their horizontal and vertical separation (correlation parameter = 0.4). The noise SD was set to 5 or 20 per voxel per run. To account for correlations in voxel noise, we estimated the noise covariance matrix from 600 independently simulated noise-only observations, which was used for noise whitening in PCM and crossnobis RSA.

**Model evaluation.** We compared regression, pattern component modeling (PCM), and RSA. For regression, we fitted separate linear models predicting each voxel's response from the candidate features, using responses averaged across the three runs. Model performance was quantified by the mean adjusted $R^2$ across all 121 voxels. For PCM, we used a single-component covariance model to evaluate whether the representational structure predicted by each candidate model was consistent with the observed multivoxel responses (Diedrichsen et al., 2011, 2018; Diedrichsen & Kriegeskorte, 2017). Unlike regression, which directly predicts voxel responses, PCM evaluates whether the covariance structure of activity patterns across stimuli matches that predicted by the feature model. After mean-centering the feature matrix X, we constructed a representational component $G = X_c X_c^T/p$, where $X_c$ is the centered feature matrix and p is the number of features. The stimulus covariance was modeled as $\Sigma = \theta G + \sigma^2 I$, where θ is a nonnegative parameter scaling the model-predicted covariance structure, and $\sigma^2$ represents the residual noise variance. We then modeled a shared stimulus-related signal with independent measurement noise across runs. Before fitting, voxel responses were whitened using the estimated noise covariance matrix. The signal and noise variance parameters were estimated by maximizing the joint likelihood across all three runs. The maximized log-likelihood served as the PCM score, with higher values indicating better model fit.

**RSA distance estimators.** For RSA, we evaluated four combinations of feature and neural distance measures: (1) correlation distance for both feature and neural RDMs, (2) Euclidean distance for both, (3) squared Euclidean distance for both, and (4) squared Euclidean distance for the feature RDM and cross-validated squared Mahalanobis distance (crossnobis) for the neural RDM (Nili et al., 2014; Walther et al., 2016). Feature RDMs were computed from the original, unstandardized feature values. Neural RDMs based on correlation, Euclidean, and squared Euclidean distances were computed from voxel responses averaged across the three runs. In contrast, crossnobis distances were estimated by combining response differences between stimuli across independent runs, weighted by the inverse of the estimated voxel-noise covariance matrix. Crossnobis neural RDMs were paired with squared Euclidean feature RDMs because crossnobis estimates squared distances in a noise-whitened neural space. For each of the four combinations, correspondence between feature and neural RDMs was quantified using both Pearson and Spearman correlations, yielding eight RSA implementations. These comparisons allowed us to examine how different distance measures influence model-selection accuracy.

## Feature reweighting of multivoxel representations

To examine whether learning feature weights improves representational correspondence and model-selection accuracy, we compared fixed RSA with voxel-encoding RSA (veRSA), following the approaches of Konkle and Alvarez (2022) and Conwell et al. (2024). We used the same multivoxel data-generating procedures and candidate models described above, but evaluated model performance on held-out stimuli using five-fold cross-validation. The three runs were first averaged to obtain a single multivoxel response pattern for each stimulus. Stimuli were then randomly divided into five equal folds. In each iteration, four folds were used for training, and the remaining fold was used for testing. This procedure was repeated until each fold had served as the test set once. Both candidate models used identical training and test splits.

**Feature reweighting.** For veRSA, we fitted separate ridge regression models for each of the 121 voxels, with the 20 stimulus features as predictors and the corresponding voxel responses as the outcome. This procedure was applied separately to the generating and rival models. Within each training fold, features were standardized, and the regularization parameter for each voxel was selected from values ranging from 0.1 to $10^7$. The optimal parameter was chosen by leave-one-out cross-validation within the training data, maximizing the Pearson correlation between observed and predicted voxel responses. The ridge models were then fitted using all training stimuli, and the resulting feature-to-voxel weights were used to predict the responses of all 121 voxels to the held-out stimuli. All standardization, parameter selection, and model fitting were based exclusively on the training data.

**Model evaluation.** For veRSA, representational similarity matrices (RSMs) were constructed from the predicted and observed multivoxel responses to the held-out stimuli, using Pearson correlations between stimulus patterns across voxels. For fixed RSA, the model RSM was constructed directly from the original, unweighted features of the same held-out stimuli, without fitting feature weights. Both approaches were evaluated against the identical observed neural RSM. Model-response correspondence was quantified as Pearson or Spearman correlation between the corresponding RSMs. For each method and candidate model, scores were averaged across the five folds, and the candidate with the higher average score was selected. We evaluated both the absolute representational correspondence of the generating and rival models and their model-selection accuracy. The primary analysis varied sample size (n = 100, 200, 300, 400, and 500) and noise level (SD = 5 and 20), with one active feature replaced (k = 1). An additional exploratory analysis at n = 200 examined k = 1 and 5 across noise SDs of 5, 10, 15, and 20, under both independent and spatially correlated noise.

### Empirical comparison of frequency and affect models

**Dataset.** To extend our method comparison to empirical behavioral data, we analyzed visual lexical-decision times and two sets of stimulus features obtained from the South Carolina Psycholinguistic Metabase (SCOPE; Gao et al., 2023), including visual lexical-decision time from the English Lexicon Project (Balota et al., 2007); nine frequency measures: Freq_SUBTLEXUS (Brysbaert & New, 2009), Freq_SUBTLEXUK (Van Heuven et al., 2014), Freq_SUBTLEXUS_Zipf (Brysbaert & New, 2009), Freq_SUBTLEXUK_Zipf (Van Heuven et al., 2014), Freq_Blog, Freq_Twitter, and Freq_News (Gimenes & New, 2016), Freq_Cob (Baayen et al., 1996), and Freq_HAL (Lund & Burgess, 1996); and nine affective measures: Valence_Warr, Arousal_Warr, and Dominance_Warr from Warriner et al. (2013), Valence_Glasgow, Arousal_Glasgow, and Dominance_Glasgow from Scott et al. (2019), and Valence_NRC, Arousal_NRC, and Dominance_NRC from Mohammad (2018). All variables were matched by word form, and words with missing values were excluded, yielding a final dataset of 4,026 words. From this dataset, we constructed two predictor sets: (1) a frequency-composite model comprising the nine frequency variables and (2) an affect-composite model comprising the nine affective variables. Both models contained substantial multicollinearity, with effective predictor ranks of seven and nine, respectively. The two models were evaluated against the same lexical-decision responses.

**Model evaluation.** We compared the frequency and affect models using regression and RSA. For each analysis, predictor variables were standardized across the included words. Regression models were fitted using ordinary least squares with an intercept, and model performance was quantified using adjusted $R^2$, accounting for the effective number of predictors. For RSA, feature RDMs were constructed using either correlation distance or Euclidean distance, whereas the response RDM was calculated from absolute differences in lexical-decision times between words. Model-response correspondence was quantified using Pearson or Spearman correlation between the corresponding RDMs, resulting in four RSA implementations. Within each method, the model with the higher score was preferred. To examine the effect of sample size on model preference, we additionally drew 1,000 random subsets without replacement at each sample size (n = 50, 100, 200, 300, 400, and 500). Both models and all methods were evaluated using the same subsets. We then calculated the proportion of subsets in which each method preferred the frequency model.

**PCA preprocessing.** As a supplementary analysis, we examined whether model preferences depended on feature preprocessing by comparing the original standardized features with PCA-rotated and PCA-whitened features. PCA was performed separately for each candidate model without using the response data. PCA rotation retained all nine nominal components without rescaling, whereas PCA whitening additionally standardized the nonzero principal components to unit variance. These analyses were conducted on both the full dataset and the randomly sampled subsets to assess how changes in feature representation influenced the relative performance of regression and RSA.

## Results

### Univariate model selection

We first compared the ability of regression and RSA to identify the generating model over a competing feature model under three data-generating schemes (**Figure 2**). In the reference condition ($n = 200$, $k = 5$, noise SD = 5), regression achieved a model-selection accuracy of 94.2% (95% Wilson CI: 92.6-95.5%) under direct-linear generation, compared with 91.6% for correlation-distance RSA using Pearson correlation and 85.3% using Spearman correlation (**Figure 2A**). Regression significantly outperformed RSA/Spearman (Holm-adjusted $p = 3.8 \times 10^{-11}$), but not RSA/Pearson ($p = 0.772$).

Under geometry-to-first-order generation with a linear readout, model-selection accuracy was 97.1% for regression, 99.0% for correlation-distance RSA/Pearson, and 98.7% for correlation-distance RSA/Spearman (**Figure 2D**). Neither RSA implementation differed significantly from regression after Holm correction ($p = 0.061$ and $0.693$, respectively). In contrast, under the sigmoid readout, RSA outperformed regression, with selection accuracies of 99.0% for RSA/Pearson and 99.3% for RSA/Spearman, compared with 95.9% for regression (**Figure 2G**; Holm-adjusted $p = 1.1 \times 10^{-4}$ and $3 \times 10^{-6}$, respectively). However, RSA performance also depended on the distance measure. For example, Euclidean-distance RSA/Pearson achieved accuracies of only 85.4% and 83.4% under the geometry-linear and geometry-sigmoid conditions, respectively.

Across all three data-generating schemes, model-selection accuracy generally increased with sample size and the number of replaced features, but decreased with increasing measurement noise (**Figure 2**). For example, at $k = 1$, regression accuracy ranged from 69.2% to 75.8% across the three schemes, increasing to 99.8–99.9% at $k = 10$. These increases also produced ceiling effects at larger sample sizes or replacement counts, limiting discrimination between methods. Comparisons away from ceiling were available at smaller sample sizes, fewer replaced features, and higher noise levels. For example, under geometry-linear generation, correlation-distance RSA/Pearson achieved 95.6% accuracy compared with 87.3% for regression at $n = 100$, $k = 5$, and noise SD = 5 (**Figure 2D**). With fewer replaced features ($n = 200$, $k = 1$, noise SD = 5), the corresponding accuracies were 83.5% and 75.8% (**Figure 2E**). At higher noise ($n = 200$, $k = 5$, noise SD = 10), they were 91.2% and 78.0% (**Figure 2F**).

Together, these results revealed distinct patterns of model-selection performance across the three generating schemes. Under direct-linear generation, regression showed an advantage over correlation-distance RSA/Spearman, consistent with its ability to directly estimate feature-to-response weights, although its advantage over RSA/Pearson was not statistically significant. Under geometry-linear generation, both regression and correlation-distance RSA achieved near-ceiling accuracy without significant differences in the reference condition. However, correlation-distance RSA outperformed regression in the smaller-sample, fewer-replacement, and higher-noise conditions described above, indicating that the near-ceiling reference comparison obscured differences evident under more challenging selection conditions. Under geometry-sigmoid generation, correlation-distance RSA significantly outperformed regression using either Pearson or Spearman correlation in the reference condition, whereas this advantage did not extend to Euclidean-distance RSA. These findings suggest that the relative model-selection performance of regression and RSA depends on how stimulus features are mapped onto observed responses, including whether responses are generated directly from features or through their representational geometry, how that geometry is transformed into responses, and which distance measure is used for RSA.

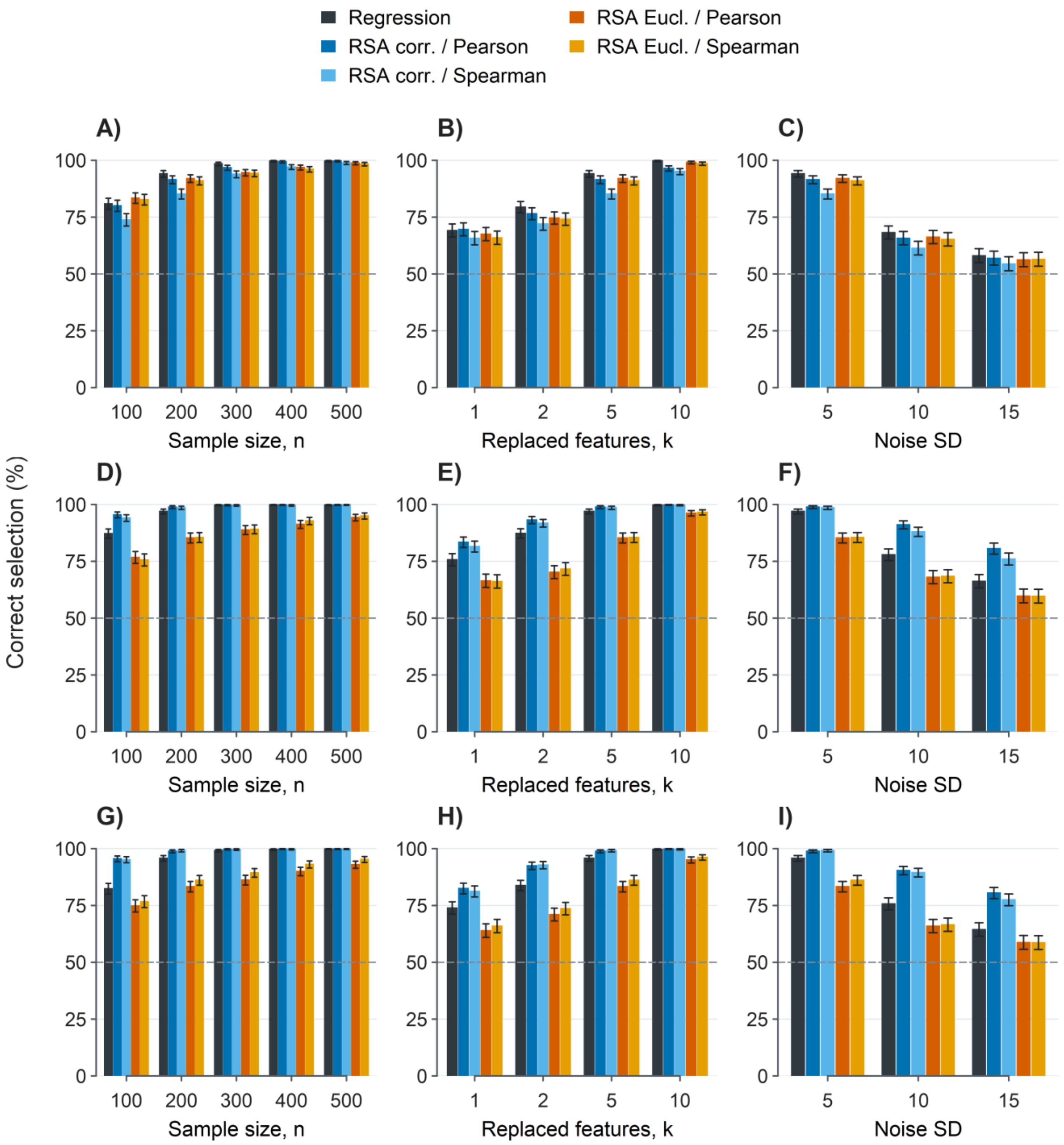

**Figure 2. Univariate model selection accuracy across data-generating schemes.** (A–C) Direct-linear generation; (D–F) geometry-to-first-order generation with a linear readout; and (G–I) geometry-to-first-order generation with a sigmoid readout. Within each scheme, panels show the effects of sample size (A, D, G), number of replaced features (B, E, H), and measurement noise (C, F, I). Bars represent the proportion of 1,000 replications in which the generating model was correctly selected, with error bars indicating 95% Wilson confidence intervals. RSA was evaluated using correlation or Euclidean feature distances, with Pearson or Spearman correlation to quantify correspondence with the response RDM. The dashed line indicates chance-level performance (50%).

## Univariate controls of dimensionality, covariance and preprocessing

We next examined how feature dimensionality, covariance structure, and preprocessing influenced model-selection performance. Increasing the number of features from 20 to 60 by adding independent, zero-coefficient features slightly reduced regression accuracy from 94.2% to 90.0%, whereas correlation-distance RSA/Pearson accuracy increased from 91.6% to 98.5% (**Figure S1**). Because the underlying signal and response were held constant, these results indicate that adding irrelevant features can affect regression and RSA differently.

We then examined the effect of feature covariance by increasing the correlation among features with zero coefficients from 0.2 to 0.8. Under direct-linear generation, regression accuracy remained relatively stable (94.2% to 93.8%), whereas correlation-distance RSA/Pearson accuracy decreased substantially (91.6% to 76.1%; **Figure S2A**). Because these features did not contribute to the generating response, this pattern indicates that correlations among irrelevant features can affect RSA-based model selection even when the underlying stimulus-response relationship remains unchanged.

Under geometry-to-first-order generation, increasing feature covariance reduced model-selection accuracy for both regression and RSA, although the reductions were larger for regression. For the geometry-linear condition, regression accuracy decreased from 97.1% to 80.9%, compared with 99.0% to 95.4% for correlation-distance RSA/Pearson (**Figure S2B**). Similarly, under the geometry-sigmoid condition, regression accuracy decreased from 95.9% to 77.0%, whereas RSA/Pearson accuracy decreased from 99.0% to 95.8% (**Figure S2C**).

Finally, we examined how PCA rotation and whitening affected model-selection accuracy (**Figure S3**). Full-component PCA rotation left regression and Euclidean-distance RSA unchanged across all three generating schemes. In contrast, it improved correlation-distance RSA under direct-linear generation but reduced its accuracy under both geometry-linear and geometry-sigmoid generation. PCA whitening further reduced model-selection accuracy for both correlation-distance and Euclidean-distance RSA across all three schemes, with particularly pronounced reductions for Euclidean-distance RSA, which approached chance-level performance under geometry-linear generation. Regression performance remained unchanged after whitening.

Together, these control analyses revealed that the number of irrelevant features and their correlation structure can have contrasting effects on model-selection performance. Under direct-linear generation, adding independent irrelevant features improved correlation-distance RSA/Pearson accuracy, whereas increasing correlations among these features substantially impaired it, despite the generating response remaining unchanged. Under geometry-based generation, stronger feature correlations reduced accuracy for both methods, but also altered the underlying response-generating geometry. The PCA controls further showed that transformations leaving regression performance unchanged can substantially alter RSA performance, particularly after feature whitening. These findings highlight that model-selection performance depends not only on feature dimensionality or collinearity, but also on how the feature space is structured and represented.

## Sensitivity to differences in signal strength

We next examined how effectively regression and RSA distinguished between independently generated datasets with different signal strengths (**Figure 3**). Under direct-linear generation, regression showed higher identification rates than correlation-distance RSA, particularly at lower noise levels (**Figure 3A-C**). At n = 200 and noise SD = 5, regression correctly identified the stronger-signal dataset in 80.7% of replications, compared with 67.3% for correlation-distance RSA/Pearson (Holm-adjusted $p = 5 \times 10^{-14}$). The corresponding independent-samples Cohen's d values were 1.24 (95% bootstrap CI: 1.15-1.34) and 0.67 (0.58-0.75), respectively (**Figure S4A**). As noise increased, identification accuracy declined and the advantage of regression became less pronounced, particularly at smaller sample sizes.

Under geometry-to-first-order generation with a linear readout, regression and correlation-distance RSA/Pearson showed similar identification performance at low noise levels (**Figure 3D–F**). At n = 200 and noise SD = 5, identification accuracy was 93.1% for regression and 92.4% for RSA/Pearson, with no significant difference after Holm correction (p = 1.000). RSA/Spearman showed a slightly lower accuracy of 89.3%, which differed significantly from regression (Holm-adjusted $p = 0.033$). Increasing noise reduced identification accuracy and generally narrowed the differences between methods.

Under geometry-to-first-order generation with a sigmoid readout, regression and RSA showed considerably lower identification accuracy than under the geometry-linear condition (**Figure 3G–I**). At $n = 200$ and noise SD = 5, regression achieved an accuracy of 60.1%, compared with 64.5% for correlation-distance RSA/Spearman, although this difference was not significant after Holm correction ($p = 0.454$). RSA/Spearman sometimes outperformed RSA/Pearson, but this pattern was not consistent across sample sizes and noise levels. The corresponding Cohen's d values are presented in **Figure S4**.

Together, these results indicate that regression was more sensitive to differences in signal strength under direct-linear generation, particularly at low noise levels, whereas its advantage was reduced under geometry-linear generation. Under geometry-sigmoid generation, neither regression nor RSA consistently demonstrated a clear advantage, and identification accuracy was substantially lower. Notably, the method differences observed in this signal-strength analysis did not fully mirror those in the common-response model-selection simulations, highlighting the distinction between detecting differences in signal strength and identifying the generating feature model.

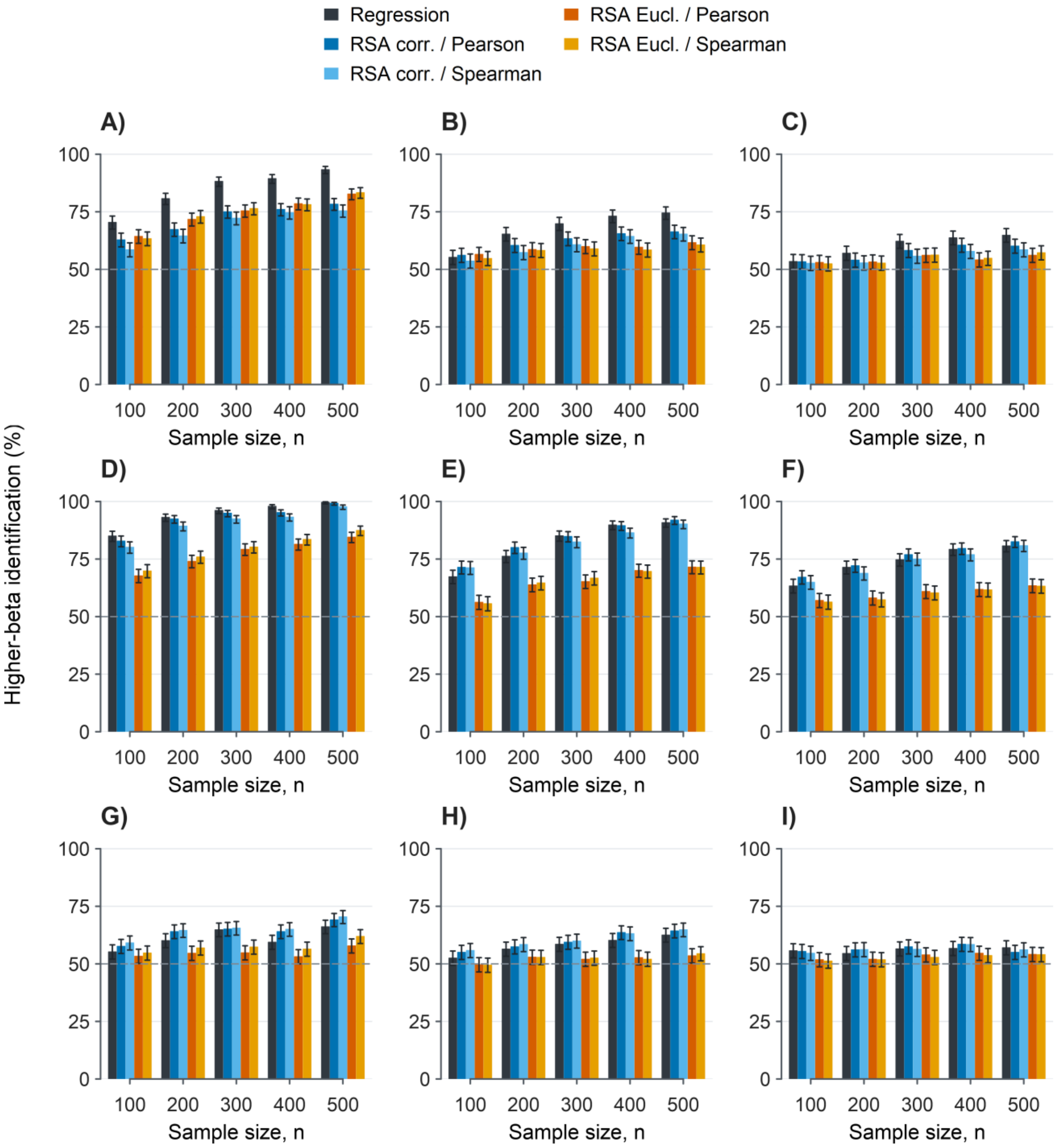

**Figure 3. Sensitivity to differences in signal strength between independently generated datasets.** (A–C) Direct-linear generation; (D–F) geometry-to-first-order generation with a linear readout; and (G–I) geometry-to-first-order generation with a sigmoid readout. Columns correspond to noise SDs of 5, 10, and 15, and the horizontal axis represents sample size (n = 100, 200, 300, 400, and 500). For each method, a successful identification was recorded when the dataset generated with $\beta = 0.5$ received a higher score than the independently generated dataset with $\beta = 0.4$. Bars show the proportion of successful identifications across 1,000 replications, with error bars indicating 95% Wilson confidence intervals. The dashed line indicates chance-level performance (50%).

## Multivoxel model selection and distance estimators

We next examined model-selection performance using simulated multivoxel responses, in which different stimulus features were mapped onto distinct spatial activity patterns (**Figure 4**). At n = 200 and per-run noise SD = 20, model-selection accuracy under independent voxel noise was 92.3% for regression, 94.7% for PCM, and 91.4% for correlation-distance RSA/Pearson (**Figure 4A**). Under spatially correlated noise, PCM accuracy increased to 98.4%, whereas regression and RSA/Pearson accuracy decreased to 86.5% and 88.2%, respectively (**Figure 4B**). In this condition, PCM significantly outperformed regression by 11.9 percentage points (95% paired-bootstrap CI: 9.9–14.0; Holm-adjusted $p = 3.7 \times 10^{-31}$). This advantage was also evident in a condition where all three methods were below ceiling: at n = 100 and per-run noise SD = 20 under spatially correlated noise, accuracy was 88.7% for PCM, 69.4% for regression, and 72.7% for correlation-distance RSA/Pearson (**Figure 4B**). PCM outperformed regression by 19.3 percentage points (95% paired-bootstrap CI: 16.4–22.3; Holm-adjusted $p < 0.001$). At larger sample sizes, particularly n = 300–500, several methods approached ceiling, limiting the ability to distinguish their performance (**Figure 4**). These results suggest that PCM can provide an advantage in model selection when voxel noise is spatially correlated.

We further examined how different distance estimators influenced RSA model-selection performance (**Figure 5**). At n = 200 and noise SD = 20, crossnobis RSA/Pearson achieved an accuracy of 62.3% under independent voxel noise, similar to the 63.2% obtained with squared Euclidean RSA/Pearson (**Figure 5A**; Holm-adjusted $p = 1.000$). Under spatially correlated noise, however, crossnobis RSA/Pearson achieved a higher accuracy of 66.5%, compared with 59.8% for squared Euclidean RSA/Pearson (**Figure 5B**), representing a significant improvement of 6.7 percentage points (95% paired-bootstrap CI: 3.5–9.8; Holm-adjusted $p = 0.009$). Thus, crossnobis improved model selection relative to squared Euclidean distance under correlated, but not independent, voxel noise. Because crossnobis incorporates both cross-validation and noise-precision weighting, the contributions of these two procedures cannot be separated in the present comparison. At the lower noise level (SD = 5), most methods achieved near-ceiling accuracy, limiting the ability to distinguish their performance. The main figures therefore focus on SD = 20, and corresponding Spearman RSA results presented in **Figures S5-S6**.

Together, these findings highlight the importance of voxel-noise structure in multivoxel model selection. PCM showed particularly strong performance under spatially correlated noise, while crossnobis provided an advantage over conventional squared Euclidean RSA under the same noise condition. These results indicate that accounting for the spatial structure of neural measurement noise can improve model discrimination, although the benefits depend on the method and distance estimator used.

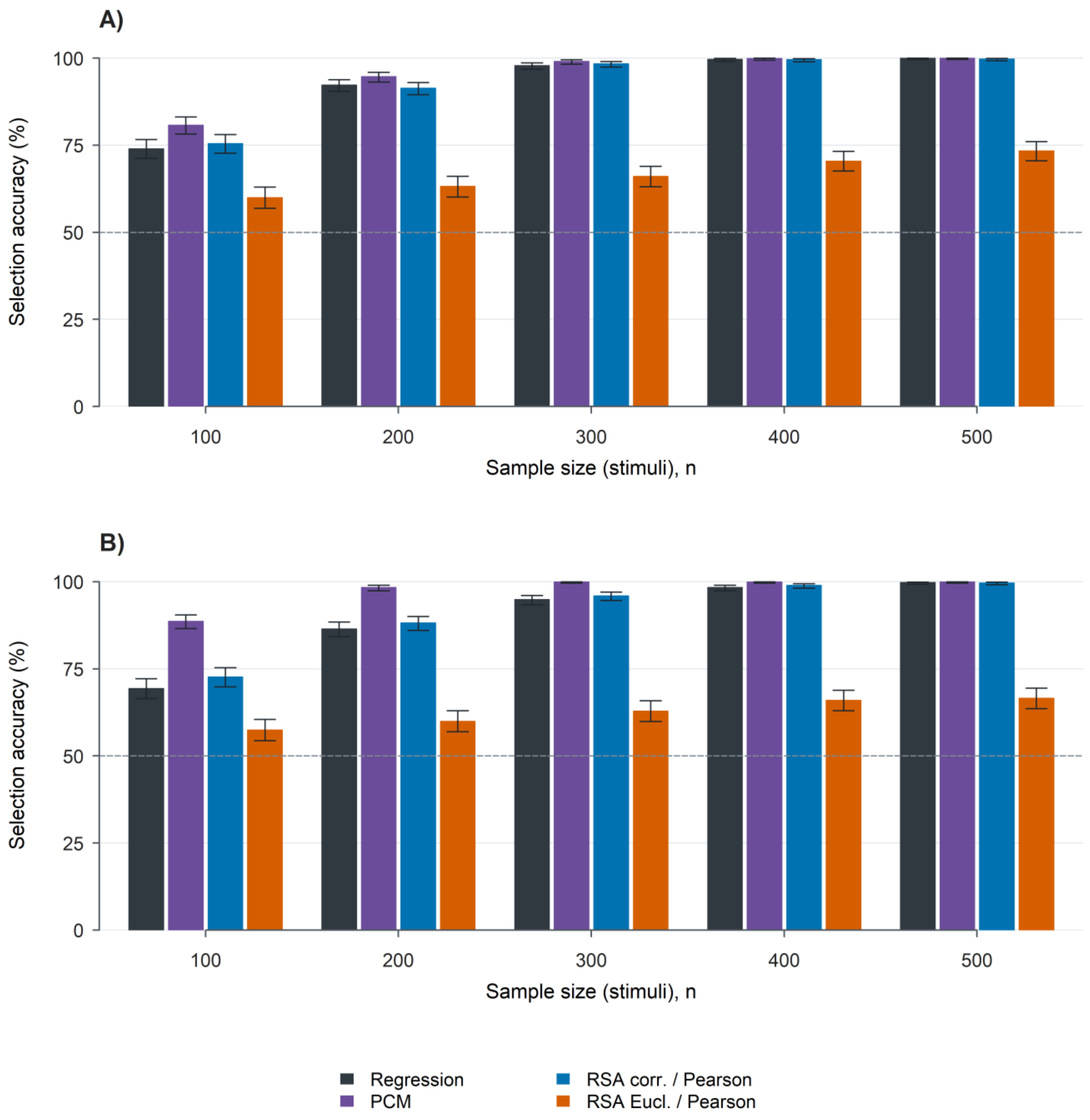

**Figure 4. Model selection with feature-specific multivoxel patterns.** (A) Independent voxel noise; (B) spatially correlated voxel noise. Multivoxel responses were generated from ten active features mapped onto distinct spatial patterns across 121 voxels and three runs. Model-selection accuracy was evaluated across sample sizes (n = 100, 200, 300, 400, and 500), with one active feature replaced (k = 1) and per-run noise SD = 20. Bars show the proportion of 1,000 replications in which the generating model was correctly selected, with error bars indicating 95% Wilson confidence intervals. The dashed line indicates chance-level performance (50%).

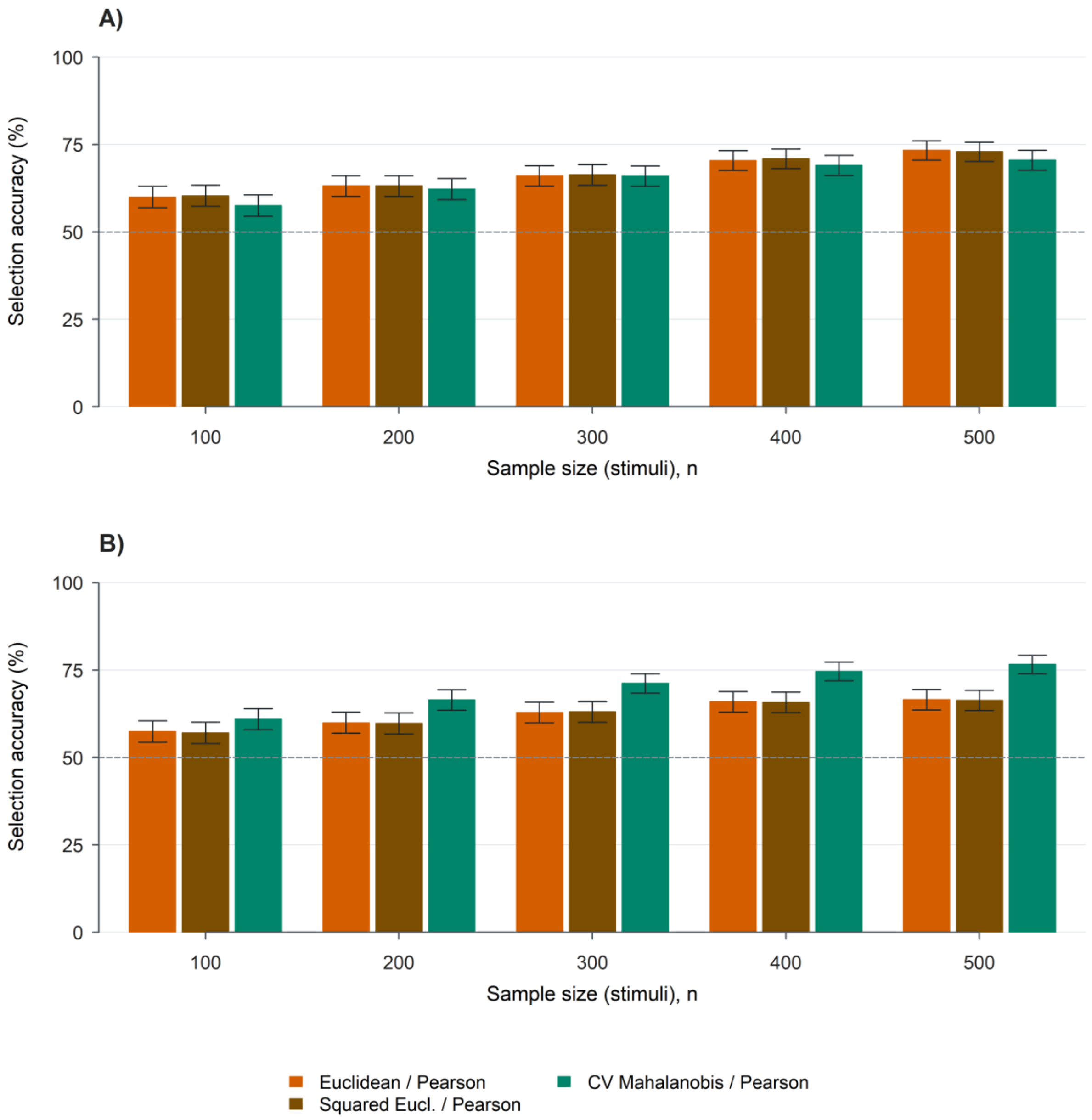

**Figure 5. Effects of neural distance estimators on multivoxel RSA model selection.** (A) Independent voxel noise; (B) spatially correlated voxel noise. RSA was evaluated using Euclidean, squared Euclidean, and cross-validated squared Mahalanobis (crossnobis) neural distances, with Pearson correlation used to quantify model-response RDM correspondence. Euclidean and squared Euclidean neural RDMs were compared with feature RDMs constructed using the corresponding distance measures, whereas crossnobis neural RDMs were compared with squared Euclidean feature RDMs. Crossnobis distances were computed using independent-run response differences weighted by estimated voxel-noise precision. The simulation settings matched those in Figure 4. Bars show model-selection accuracy across 1,000 replications, with error bars indicating 95% Wilson confidence intervals. The dashed line indicates chance-level performance (50%).

## Feature reweighting of multivoxel representations

We next examined whether learning feature-to-voxel weights improved held-out representational correspondence and model-selection accuracy. Compared with fixed RSA, voxel-encoding RSA (veRSA) generally improved representational correspondence for both the generating and rival models, particularly at lower noise levels (**Figure 6**). At n = 200 and noise SD = 5 under independent voxel noise, the generating model's Pearson RSM correlation increased from 0.367 with fixed RSA to 0.616 with veRSA, while the rival model's correlation also increased from 0.333 to 0.585 (**Figure 6A**). At the higher noise level (SD = 20), the improvements were substantially smaller, with correlations increasing from 0.039 to 0.044 for the generating model and from 0.036 to 0.040 for the rival model (**Figure 6C**). At n = 100 under high-noise conditions, feature reweighting did not consistently improve representational correspondence, suggesting that its benefits depended on the amount of noise and the available training data.

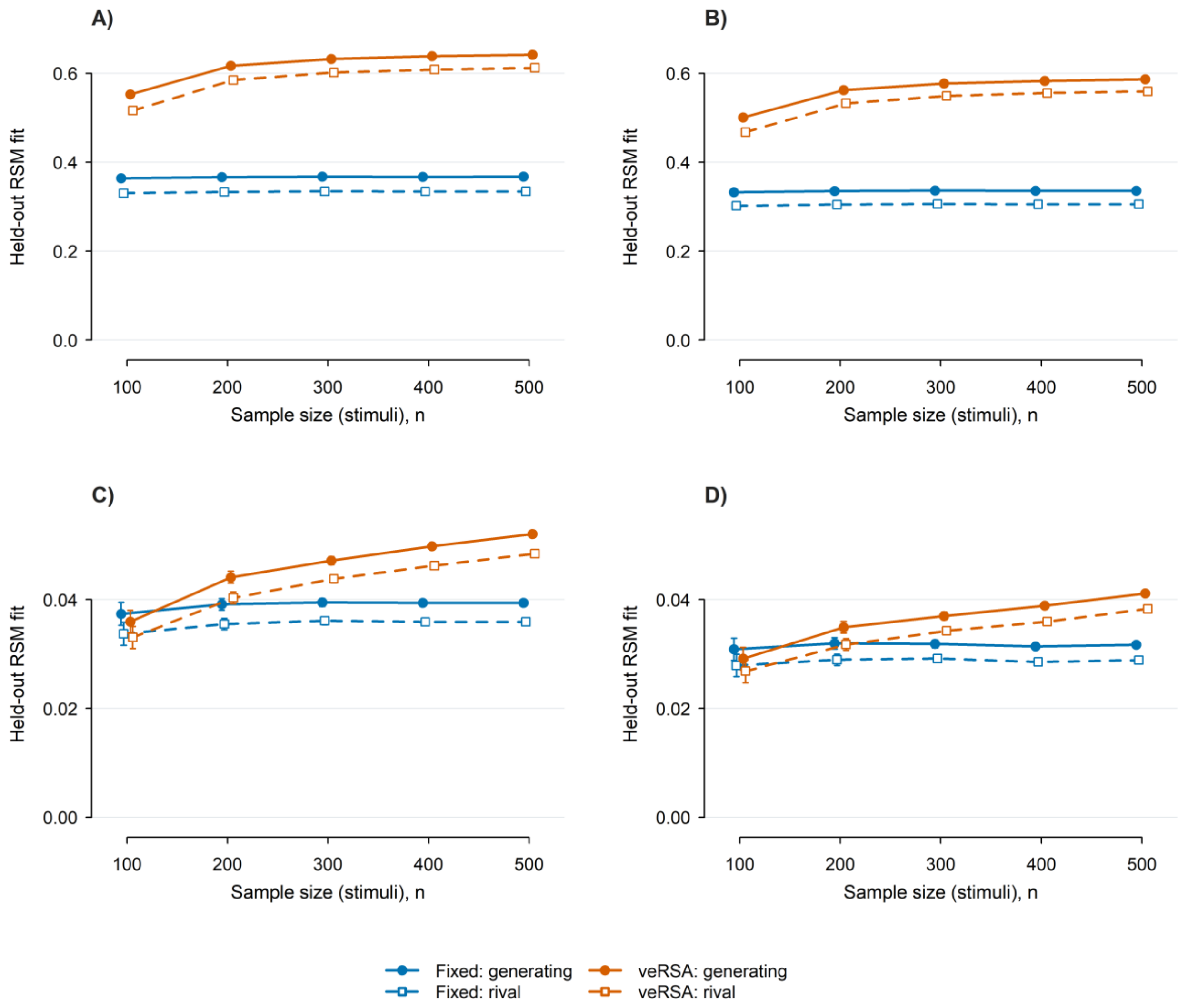

**Figure 6. Effects of feature reweighting on held-out representational correspondence.** (A) Independent voxel noise, SD = 5; (B) spatially correlated voxel noise, SD = 5; (C) independent voxel noise, SD = 20; and (D) spatially correlated voxel noise, SD = 20. Fixed RSA and voxel-encoding RSA (veRSA) were evaluated for both the generating and rival models using the same held-out stimuli. Representational correspondence was quantified using Pearson correlations between model-predicted and observed RSMs, averaged across five cross-validation folds. Points represent mean correlations across 1,000 replications, with error bars indicating 95% replication-bootstrap confidence intervals.

Despite improving representational correspondence, feature reweighting did not consistently improve model-selection accuracy (**Figure 7**). At n = 200 and noise SD = 20, fixed RSA and veRSA achieved selection accuracies of 75.0% and 72.2%, respectively, under independent voxel noise (**Figure 7A**). Under spatially correlated noise, the corresponding accuracies were 72.8% and 67.6% (**Figure 7B**), with veRSA performing significantly worse than fixed RSA (difference = −5.2 percentage points, 95% paired-bootstrap CI: −8.2 to −2.2; Holm-adjusted $p = 0.033$). At the lower noise level (SD = 5), both methods frequently approached ceiling performance. The corresponding Spearman results for representational correspondence and model-selection accuracy are presented in **Figures S7 and S8**, respectively. Additional exploratory analyses varying noise level and the number of replaced features found no significant improvement in model-selection accuracy for veRSA over fixed RSA after Holm correction (**Figures S9-S12**).

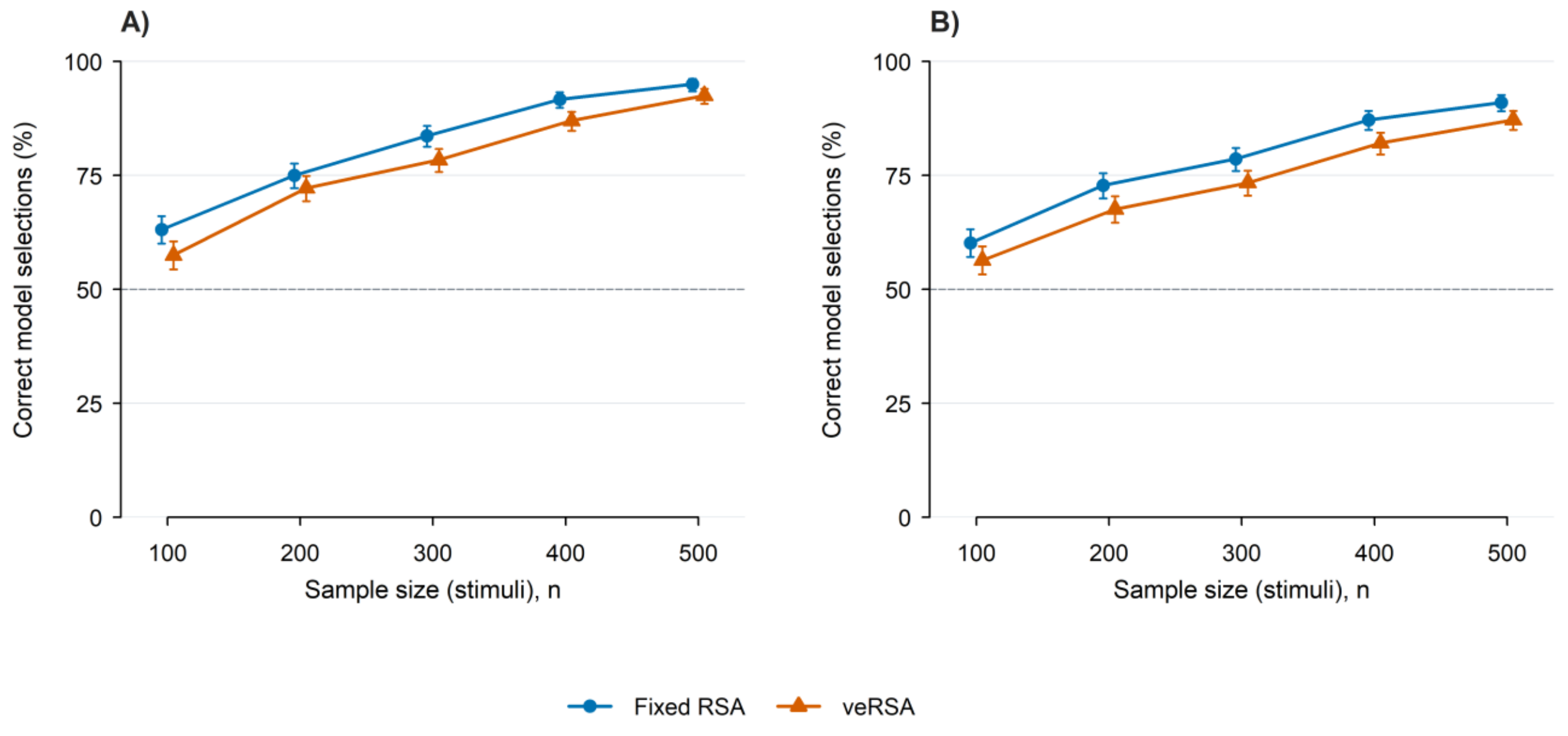

**Figure 7. Effects of feature reweighting on held-out model-selection accuracy.** (A) Independent voxel noise; (B) spatially correlated voxel noise. Fixed RSA and veRSA were compared at noise SD = 20, with one active feature replaced ($k = 1$). Both methods were evaluated using identical held-out stimulus folds, and the generating model was selected when its average RSM correlation across folds exceeded that of the rival model. Points show the proportion of correct selections across 1,000 replications, with error bars indicating 95% Wilson confidence intervals. The dashed line indicates chance-level performance (50%).

Together, these results demonstrate that improved representational correspondence does not necessarily translate into more accurate model selection. Although learning feature-to-voxel weights often improved the fit of the generating model, it also improved the fit of the rival model, limiting gains in discrimination between them. Thus, the benefits of feature reweighting for explaining neural representations should be distinguished from its ability to identify the generating model.

## Empirical comparison of frequency and affect models

Finally, we examined how regression and RSA distinguished between two theoretically motivated feature models using empirical lexical-decision data. Based on previous findings establishing word frequency as a strong predictor of lexical-decision times (Brysbaert & New, 2009; Brysbaert et al., 2018), we expected the frequency model to outperform the affect model. Consistent with this expectation, all methods favored the frequency model across the full dataset of 4,026 words (**Figure 8A-B**). Regression yielded an adjusted $R^2$ of 0.4172 (95% bootstrap CI: 0.3962–0.4393) for the frequency model, compared with 0.0430 for the affect model. For correlation-distance RSA/Pearson, model-response correspondence was 0.0705 for frequency and 0.0287 for affect, whereas Euclidean-distance RSA/Pearson yielded correlations of 0.3160 and 0.0156, respectively.

Although all methods agreed in their preference for the frequency model, they differed in the stability of this preference across smaller word samples (**Figure 8C**). At n = 50, regression preferred the frequency model in 99.0% of subsets, compared with 72.8% for correlation-distance RSA/Pearson and 98.2% for Euclidean-distance RSA/Pearson. As sample size increased, model preference became more consistent across methods. At n = 500, the corresponding preference rates reached 100%, 99.0%, and 100%, respectively. Thus, regression and Euclidean-distance RSA/Pearson showed more stable model preferences than correlation-distance RSA/Pearson, particularly at smaller sample sizes.

We further examined how PCA preprocessing influenced model preference. At n = 50, PCA rotation increased the frequency-model preference rate for correlation-distance RSA/Pearson from 72.8% to 99.0%, while PCA whitening yielded a rate of 97.9%. In contrast, Euclidean-distance RSA/Pearson was unaffected by PCA rotation but showed reduced preference stability after whitening, with its frequency-model preference rate declining from 98.2% to 86.8%. Neither transformation reversed the preference for the frequency model in the full dataset.

Together, these empirical findings were broadly consistent with the direct-linear simulations. First, regression showed more stable model selection than correlation-distance RSA at smaller sample sizes, while Euclidean-distance RSA also demonstrated high stability. Second, PCA rotation left regression and Euclidean-distance RSA unchanged while improving correlation-distance RSA, consistent with the direct-linear simulation results. Third, PCA whitening reduced Euclidean-distance RSA performance in both simulated and empirical data. These converging findings suggest that regression provides relatively stable model discrimination under conditions resembling direct-linear mappings, whereas RSA-based model selection is more sensitive to the choice of distance measure and feature preprocessing.

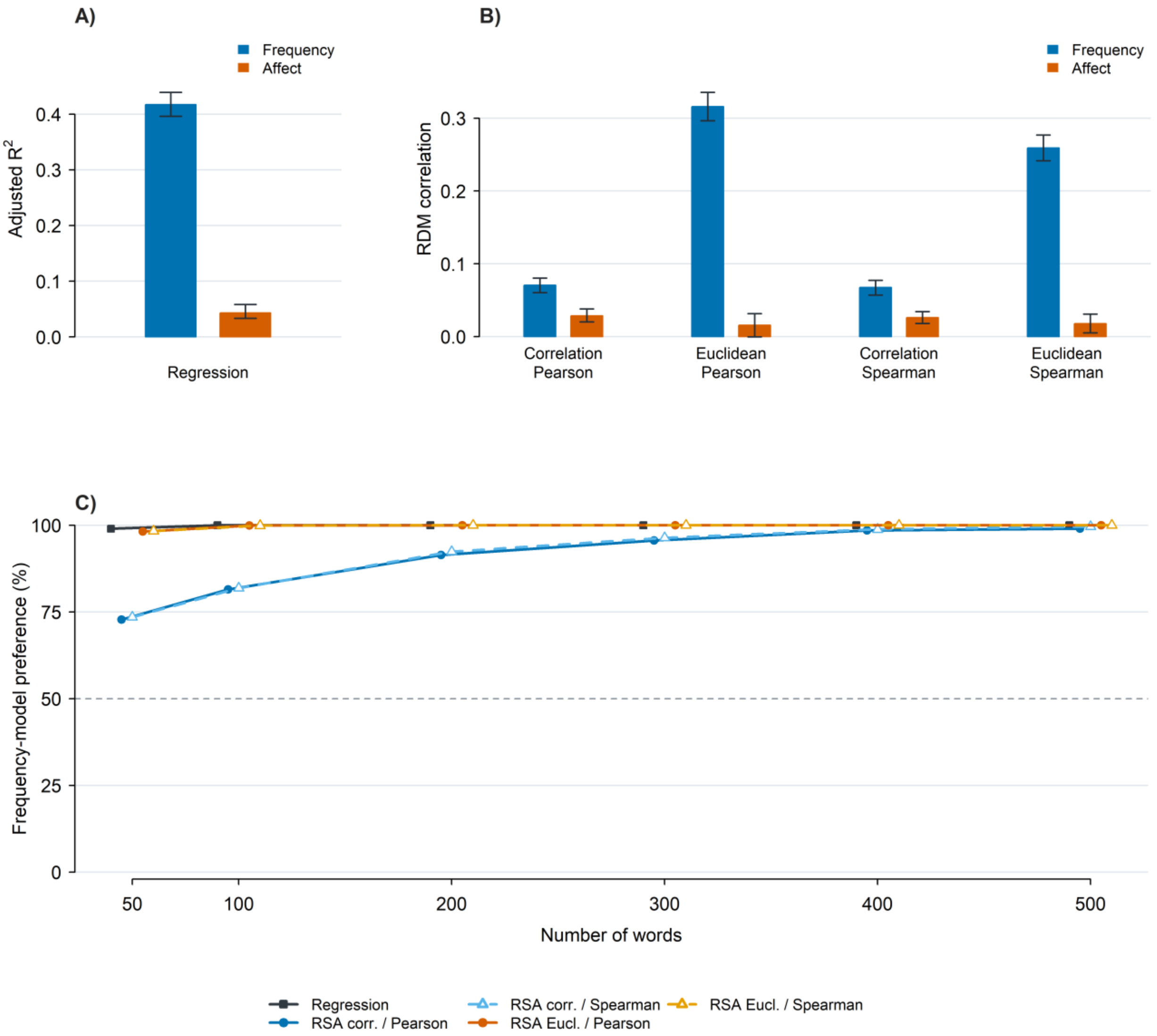

**Figure 8. Empirical comparison of frequency and affect feature models.** (A) Regression adjusted $R^2$ for the full dataset; (B) RSA model-response correspondence using different feature distance measures and RDM correlation metrics; and (C) frequency-model preference rates across sample sizes. Both feature models were evaluated against the same aggregate lexical-decision responses for 4,026 English words. Preference rates were calculated across 1,000 matched random subsets at each sample size and reflect the stability of model preference rather than known-truth model-selection accuracy.

## Discussion

In the present study, we examined when regression, representational similarity analysis (RSA), and pattern component modeling (PCM) differ in their ability to distinguish competing feature models. Across simulations with known generating models, no single approach consistently achieved the highest model-selection accuracy. Regression showed an advantage over correlation-distance RSA/Spearman under direct-linear generation, whereas correlation-distance RSA outperformed regression under geometry-sigmoid generation and, under more challenging selection conditions, geometry-linear generation. These differences also depended on the choice of representational distance. Additional analyses revealed the influence of feature-space structure, preprocessing, and spatial measurement noise on model selection. Furthermore, feature reweighting improved representational correspondence without consistently improving model-selection accuracy. Together, these findings indicate that method performance depends on the assumed stimulus-response mapping, the representation and measurement of the data, and the specific objective of model evaluation.

A key finding is that the manner in which stimulus features map onto observed responses shapes model-selection performance. When data were generated via a direct-linear scheme, regression profited from its capacity to estimate feature-to-response weights, in line with the linear mappings widely assumed in neural encoding models (Kay et al., 2008; Huth et al., 2016). Nevertheless, this advantage was significant compared with correlation-distance RSA/Spearman but not RSA/Pearson. A different picture emerged when responses were instead generated from representational geometry, a framework inspired by similarity-based theories of categorization (Nosofsky, 1986). Under geometry-linear generation, both regression and correlation-distance RSA reached near-ceiling accuracy in the reference condition, with no significant differences between them. This comparison had limited ability to distinguish the methods and does not establish equivalence; correlation-distance RSA showed an advantage at smaller sample sizes, with fewer replaced features, and at higher noise levels (**Figure 2D–F**). Under geometry-sigmoid generation, correlation-distance RSA outperformed regression, whereas Euclidean-distance RSA did not show the same benefit. Thus, model-selection performance depends not simply on whether responses originate from representational geometry, but also on how that geometry is transformed into responses and how representational distances are defined.

Importantly, differences between regression and RSA cannot be attributed solely to the transformation of first-order data into second-order dissimilarities. Regression estimates feature-to-response weights, whereas fixed RSA evaluates predefined representational geometries without fitting feature contributions. Moreover, different distance measures preserve different properties of the feature space. Correlation distance removes differences in overall pattern level and magnitude, whereas Euclidean distance retains them (Kriegeskorte, Mur, & Bandettini, 2008; Nili et al., 2014). Therefore, differences in model-selection performance may reflect both feature-weight estimation (Jozwik et al., 2016; Kaniuth & Hebart, 2022) and representational distance, rather than information loss through second-order transformation alone. This interpretation is consistent with the common representational modeling framework, which views encoding models, RSA, and PCM as complementary approaches to characterizing representational structure (Diedrichsen & Kriegeskorte, 2017).

Additional control analyses revealed that the structure of the feature space can shape model-selection performance even when the underlying stimulus-response relationship stays the same. When data were generated under a direct-linear scheme, incorporating independent features with zero coefficients enhanced the accuracy of correlation-distance RSA while marginally diminishing regression accuracy. Conversely, raising the correlations among these irrelevant features markedly lowered RSA accuracy without materially affecting regression. These contrasting effects suggest that RSA performance depends not simply on the number of features, but also on how they are correlated and represented. Because fixed RSA evaluates distances across all supplied features, even features unrelated to the generating response can influence the resulting representational geometry (Xie et al., 2025). The PCA results further highlighted the importance of feature representation. Although PCA rotation had no effect on regression or Euclidean-distance RSA, its impact on correlation-distance RSA varied across the generating schemes. Feature whitening, meanwhile, diminished RSA accuracy in all three schemes without altering regression performance. Taken together, these results imply that PCA should not be treated as a universal remedy for multicollinearity in representational analyses. Instead, feature-space transformations ought to be assessed by considering how they modify representational geometry and, consequently, model-selection outcomes (Oswal et al., 2016).

The multivoxel simulations additionally underscored how critical the structure of measurement noise can be. When voxel noise was spatially correlated, PCM outperformed regression and correlation-distance RSA by a wide margin, aligning with its covariance-based modeling of stimulus representations and its explicit incorporation of voxel-noise covariance (Diedrichsen et al., 2011, 2018; Diedrichsen & Kriegeskorte, 2017). In a similar vein, crossnobis yielded better model-selection accuracy than squared Euclidean RSA under correlated noise, though this benefit did not emerge under independent noise. Crossnobis integrates cross-validated distance estimation with precision weighting based on voxel noise, thereby offering a way to mitigate noise-related biases in multivoxel representations (Nili et al., 2014; Walther et al., 2016). While the present comparison cannot disentangle the individual contributions of these mechanisms, these results demonstrate that accounting for neural measurement noise can be important for reliable model discrimination.

Another important finding is that improved representational correspondence does not necessarily lead to more accurate model selection. Previous work has shown that learning feature weights can improve model-brain correspondence (Jozwik et al., 2016; Kaniuth & Hebart, 2022; Konkle & Alvarez, 2022; Conwell et al., 2024). Consistent with these findings, voxel-encoding RSA (veRSA) generally improved held-out correspondence for the generating model. However, it also improved correspondence for the rival model, which retained substantial response-relevant information through shared and correlated features. Consequently, better absolute fit did not translate into improved model discrimination, and veRSA even performed worse than fixed RSA under one correlated-noise condition. This does not imply that feature reweighting is generally detrimental to model selection. Rather, its benefits may depend on the similarity between competing models, measurement noise, and available training data. Critically, model selection depends not simply on improving model correspondence, but on whether the improvement increases discrimination between competing models.

Signal-strength sensitivity and model selection also address distinct inferential objectives. In neuroimaging research, for example, the former concerns whether the same computational model shows stronger correspondence across independent fMRI datasets, whereas the latter asks which competing model better explains the same neural responses. Our simulations formalized this distinction by comparing independently generated high- and low-signal datasets versus competing feature models evaluated against a common response. The different method rankings observed across these analyses indicate that sensitivity to changes in model-response strength does not necessarily imply greater ability to discriminate competing models. Thus, stronger model-response correspondence should not automatically be interpreted as stronger evidence for one theoretical model over its alternatives (Maxwell et al., 2018; Diedrichsen & Kriegeskorte, 2017).

The empirical lexical-decision analysis provided complementary evidence for model selection using real behavioral data. Consistent with the established word-frequency effect (Brysbaert & New, 2009; Brysbaert et al., 2018), all methods favored the frequency model over the affect model, but regression and Euclidean-distance RSA showed more stable model preferences at smaller sample sizes than correlation-distance RSA. This may reflect the importance of overall word-frequency magnitude, which is preserved by Euclidean distance but removed by correlation distance (Kriegeskorte, Mur, & Bandettini, 2008). Several preprocessing effects also paralleled the direct-linear simulations: PCA rotation improved correlation-distance RSA while leaving regression and Euclidean-distance RSA unchanged, whereas whitening reduced Euclidean-distance RSA performance. These findings provide converging evidence that distance choice and feature preprocessing influence model discrimination, although empirical model preference cannot establish known-truth model-selection accuracy.

A few limitations of the current work warrant acknowledgment. First, the nonlinear readout used in our simulations relied on a single sigmoid transformation, and neither the form nor the degree of nonlinearity was systematically varied. The observed differences in model-selection performance should therefore be understood as particular to the transformation we tested, rather than as a general account of nonlinear mappings. Future research could examine whether these findings generalize to other monotonic and non-monotonic transformations, and how the relative performance of regression and RSA changes with the degree of nonlinearity. Second, although we assessed several RSA distance measures and feature reweighting, our multivoxel analyses made use of a simplified single-component PCM implementation and a single ridge-based feature-reweighting method. Future studies might explore whether more flexible representational models and alternative feature-fitting procedures produce comparable patterns of model-selection performance. Third, near-ceiling accuracy in some univariate and multivoxel conditions compressed observable differences between methods. Nonsignificant comparisons in these conditions should not be interpreted as evidence of equivalent performance. The non-ceiling conditions already included in our parameter ranges provide more informative comparisons, although the observed method differences remain specific to those simulation settings.

The current findings carry practical implications for selecting analytical methods. Regression is particularly appropriate for assessing direct feature-to-response mappings, especially when conditions are roughly linear (Naselaris et al., 2011; Kriegeskorte & Douglas, 2019). RSA continues to be useful when the research question centers on representational geometry or when no direct mapping between feature and response dimensions is available (Kriegeskorte & Kievit, 2013; Nili et al., 2014). That said, distance measures and feature preprocessing warrant careful consideration, since they determine the representational structure under evaluation. For multivoxel data, PCM and cross-validated distance estimators provide further options for modeling representational covariance and measurement noise (Diedrichsen & Kriegeskorte, 2017; Walther et al., 2016). When competing models are being compared, researchers ought to assess relative model performance directly, employ held-out data when fitting feature weights, and keep model selection distinct from comparisons of signal strength. Overall, our findings suggest that regression, RSA, and PCM should be viewed as complementary approaches rather than universally competing alternatives. Their effectiveness in model selection depends on the assumptions linking stimulus features to observed responses, the definition of representational geometry, the treatment of measurement noise, and the specific inferential objective.